\documentclass[twoside]{article}%
\usepackage{amssymb}
\usepackage{amsfonts}
\usepackage{amsmath}
\usepackage{graphicx}%
\providecommand{\U}[1]{\protect\rule{.1in}{.1in}}
\newtheorem{theorem}{Theorem}

\newtheorem{lemma}[theorem]{Lemma}

\newtheorem{remark}[theorem]{Remark}

\newenvironment{proof}[1][Proof]{\noindent\textbf{#1.} }{\ \rule{0.5em}{0.5em}}
\begin{document}

\title{\textbf{Some Exact Blowup Solutions to the Pressureless Euler Equations in
}$R^{N}$}
\author{Y\textsc{uen} M\textsc{anwai\thanks{E-mail address: nevetsyuen@hotmail.com }}\\\textit{Department of Applied Mathematics, The Hong Kong Polytechnic
University,}\\\textit{Hung Hom, Kowloon, Hong Kong}}
\date{Revised 07-Oct-2009}
\maketitle

\begin{abstract}
The pressureless Euler equations can be used as simple models of cosmology or
plasma physics. In this paper, we construct the exact solutions in non-radial
symmetry to the pressureless Euler equations in $R^{N}:$%
\begin{equation}
\left\{
\begin{array}
[c]{c}%
\rho(t,\vec{x})=\frac{f\left(  \frac{1}{a(t)^{s}}\underset{i=1}{\overset
{N}{\sum}}x_{i}^{s}\right)  }{a(t)^{N}}\text{, }\vec{u}(t,\vec{x}%
)=\frac{\overset{\cdot}{a}(t)}{a(t)}\vec{x},\\
a(t)=a_{1}+a_{2}t,
\end{array}
\right.  \label{eq234}%
\end{equation}
where the arbitrary function $f\geq0$ and $f\in C^{1};$ $s\geq1$, $a_{1}>0$
and $a_{2}$ are constants$.$\newline In particular, for $a_{2}<0$, the
solutions blow up on the finite time $T=-a_{1}/a_{2}$.

Moreover, the functions (\ref{eq234}) are also the solutions to the
pressureless Navier-Stokes equations.


\end{abstract}

\section{Introduction}

The $N$-dimensional Euler equations can be formulated in the following form:%
\begin{equation}
\left\{
\begin{array}
[c]{rl}%
{\normalsize \rho}_{t}{\normalsize +\nabla\cdot(\rho\vec{u})} &
{\normalsize =}{\normalsize 0,}\\
\rho\left[  \vec{u}_{t}+\left(  \vec{u}\cdot\nabla\right)  \vec{u}\right]
+\delta\nabla P(\rho) & {\normalsize =}0,
\end{array}
\right.  \label{eq1}%
\end{equation}
with $\vec{x}=(x_{1},x_{2},...,x_{N})\in R^{N}$ and As usual, $\rho
=\rho(t,\vec{x})$ and $\vec{u}(t,\vec{x})\in R^{N}$ are the density and the
velocity respectively. If $\delta=1$, the system is with pressure and
$P=P(\rho)$ is the pressure. The $\gamma$-law on the pressure, i.e.
\begin{equation}
P(\rho)=K\rho^{\gamma}, \label{eq2}%
\end{equation}
with $K>0$, is a universal hypothesis. If $\delta=0$, the system is
pressureless:%
\begin{equation}
\left\{
\begin{array}
[c]{rl}%
{\normalsize \rho}_{t}{\normalsize +\nabla\cdot(\rho\vec{u})} &
{\normalsize =}{\normalsize 0,}\\
\rho\left[  \vec{u}_{t}+\left(  \vec{u}\cdot\nabla\right)  \vec{u}\right]  &
{\normalsize =}0.
\end{array}
\right.  \label{pressurelessEuler}%
\end{equation}
The system can be used as models of cosmology \cite{Z} or plasma physics
\cite{BB}. There are also intensive studies for the pressureless Euler
equations (\ref{pressurelessEuler}) in the recent literature \cite{BBT},
\cite{CKR}, \cite{DD} and \cite{Mo}. The constant $\gamma=c_{P}/c_{v}\geq1$,
where $c_{p}$ and $c_{v}$ are the specific heats per unit mass under constant
pressure and constant volume respectively, is the ratio of the specific heats.
In particular, the fluid is called isothermal if $\gamma=1$. The Euler
equations (\ref{eq1}) govern the evolutionary phenomena of classical fluid
dynamics. For the detailed studies of the Euler equations (\ref{eq1}), see
\cite{CW}, \cite{L} and \cite{Ni}.

For the Euler equations (\ref{eq1}) in radial symmetry
\begin{equation}
\rho(t,x)=\rho(t,r)\text{ and }\vec{u}=\frac{\vec{x}}{r}V(t,r):=\frac{\vec{x}%
}{r}V,
\end{equation}
with $r=\left(  \sum_{i=1}^{N}x_{i}^{2}\right)  ^{1/2}$,\newline there exists
a family of solutions \cite{Li} for $\gamma>1$,%
\begin{equation}
\left\{
\begin{array}
[c]{c}%
\rho(t,r)=\left\{
\begin{array}
[c]{cc}%
\frac{^{y(r/a(t))^{1/(\gamma-1)}}}{a(t)^{N}}, & \text{ for }y(\frac{r}%
{a(t)})\geq0;\\
0, & \text{for }y(\frac{r}{a(t)})<0
\end{array}
\right.  ,\text{ }V(t,r)=\frac{\overset{\cdot}{a}(t)}{a(t)}r,\\
\overset{\cdot\cdot}{a}(t)=\frac{-\lambda}{a(t)^{^{1+N(\gamma-1)}}},\text{
}a(0)=a_{0}>0,\text{ }\overset{\cdot}{a}(0)=a_{1},\\
y(x)=\frac{(\gamma-1)\lambda}{2\gamma K}x^{2}+\alpha^{\theta-1},
\end{array}
\right.  ;
\end{equation}
\cite{Y2} for $\gamma=1,$%
\begin{equation}
\left\{
\begin{array}
[c]{c}%
\rho(t,r)=\frac{e^{y(r/a(t))}}{a(t)^{N}},u(t,r)=\frac{\overset{\cdot}{a}%
(t)}{a(t)}r\\
\overset{\cdot\cdot}{a}(t)=\frac{-\lambda}{a(t)},a(0)=a_{0}>0,\overset{\cdot
}{a}(0)=a_{1},\\
y(x)=\frac{\lambda}{2K}x^{2}+\alpha,
\end{array}
\right.
\end{equation}
where $\lambda,$ $\alpha,$ $a_{0}$ and $a_{1}$ are constants.

The analytical solutions are found because of the techniques of separation
method of self-similar solutions. The method were used to handle other similar
systems in \cite{DXY}, \cite{Li}, \cite{M1}, \cite{Y}, \cite{Y1}, \cite{Y2},
\cite{YY}, \cite{Y3}, \cite{Y4} and \cite{Y5}.

It is very natural to see that for the pressureless Euler equations
(\ref{pressurelessEuler}), there exists a class of solutions
\begin{equation}
\rho(t,r)=\frac{^{f(r/a(t))}}{a(t)^{N}},\text{ }V(t,r)=\frac{\overset{\cdot
}{a}(t)}{a(t)}r,
\end{equation}
with the arbitrary function $f\geq0$ and $f\in C^{1};$ and $a(t)>0$ and
$a(t)\in C^{1}.$\newline In this article, we have obtained the more general
results about the pressureless Euler equations (\ref{pressurelessEuler}) in
the following theorem:

\begin{theorem}
\label{thm:1}For the $N$-dimensional pressureless Euler equations
(\ref{pressurelessEuler}), there exists a family of solutions:%
\begin{equation}
\left\{
\begin{array}
[c]{c}%
\rho(t,\vec{x})=\frac{f\left(  \frac{1}{a(t)^{s}}\underset{i=1}{\overset
{N}{\sum}}x_{i}^{s}\right)  }{a(t)^{N}}\text{, }\vec{u}(t,\vec{x}%
)=\frac{\overset{\cdot}{a}(t)}{a(t)}\vec{x},\\
a(t)=a_{1}+a_{2}t,
\end{array}
\right.  \label{Yuensol}%
\end{equation}
with the arbitrary function $f\geq0$ and $f\in C^{1};$ and $s\geq1$, $a_{1}>0$
and $a_{2}$ are constants$.$\newline In particular, for $a_{2}<0$, the
solutions blow up on the finite time $T=-a_{1}/a_{2}$.
\end{theorem}

\section{Separation Method}

Regards to the continuity equation of mass (\ref{pressurelessEuler})$_{1}$, we
found that the following solution structures fit it well:

\begin{lemma}
\label{lem:generalsolutionformasseq}For the equation of conservation of mass,
\begin{equation}
\rho_{t}+\nabla\cdot\left(  \rho\vec{u}\right)  =0,
\label{massequationspherical}%
\end{equation}
there exist general solutions,%
\begin{equation}
\rho(t,\vec{x})=\frac{f\left(  \frac{1}{a(t)^{s}}\underset{i=1}{\overset
{N}{\sum}}x_{i}^{s}\right)  }{a(t)^{N}},{\normalsize \vec{u}(t,\vec{x})}%
=\frac{\dot{a}(t)}{a(t)}\vec{x}, \label{generalsolutionformassequation}%
\end{equation}
with the arbitrary function $f\geq0$ and $f\in C^{1}$; $a(t)>0$ and $a(t)\in
C^{1}$; and the constant $s\geq1$ $.$
\end{lemma}

\begin{proof}
We just plug (\ref{generalsolutionformassequation}) into
(\ref{massequationspherical}). Then, we have:%
\begin{align}
&  \rho_{t}+\nabla\bullet\vec{u}\rho+\nabla\rho\bullet\vec{u}\\
&  =\frac{\partial}{\partial t}\left[  \frac{f\left(  \frac{1}{a(t)^{s}%
}\underset{i=1}{\overset{N}{\sum}}x_{i}^{s}\right)  }{a(t)^{N}}\right]
+\left[  \nabla\bullet\frac{\dot{a}(t)}{a(t)}\vec{x}\right]  \frac{f\left(
\frac{1}{a(t)^{s}}\underset{i=1}{\overset{N}{\sum}}x_{i}^{s}\right)
}{a(t)^{N}}+\nabla\frac{f\left(  \frac{1}{a(t)^{s}}\underset{i=1}{\overset
{N}{\sum}}x_{i}^{s}\right)  }{a(t)^{N}}\bullet\frac{\dot{a}(t)}{a(t)}\vec{x}\\
&  =\frac{-N\dot{a}(t)}{a(t)^{N+1}}f\left(  \frac{1}{a(t)^{s}}\underset
{i=1}{\overset{N}{\sum}}x_{i}^{s}\right)  +\frac{1}{a(t)^{N}}\frac{\partial
}{\partial t}f\left(  \frac{1}{a(t)^{s}}\underset{i=1}{\overset{N}{\sum}}%
x_{i}^{s}\right) \\
&  +\underset{i=1}{\overset{N}{\sum}}\frac{\dot{a}(t)}{a(t)}\frac{f\left(
\frac{1}{a(t)^{s}}\underset{i=1}{\overset{N}{\sum}}x_{i}^{s}\right)
}{a(t)^{N}}+\underset{i=1}{\overset{N}{\sum}}\frac{\partial}{\partial x_{i}%
}\left[  \frac{f\left(  \frac{1}{a(t)^{s}}\underset{i=1}{\overset{N}{\sum}%
}x_{i}^{s}\right)  }{a(t)^{N}}\right]  \frac{\dot{a}(t)}{a(t)}x_{i}\\
&  =\frac{-N\dot{a}(t)}{a(t)^{N+1}}f\left(  \frac{1}{a(t)^{s}}\underset
{i=1}{\overset{N}{\sum}}x_{i}^{s}\right)  -\frac{\dot{f}\left(  \frac
{1}{a(t)^{s}}\underset{i=1}{\overset{N}{\sum}}x_{i}^{s}\right)  }{a(t)^{N}%
}\frac{s\underset{i=1}{\overset{N}{\sum}}x_{i}^{s}\dot{a}(t)}{a(t)^{s+1}}\\
&  +N\frac{\dot{a}(t)}{a(t)}\frac{f\left(  \frac{1}{a(t)^{s}}\underset
{i=1}{\overset{N}{\sum}}x_{i}^{s}\right)  }{a(t)^{N}}+\underset{i=1}%
{\overset{N}{\sum}}\frac{\dot{f}\left(  \frac{1}{a(t)^{s}}\underset
{i=1}{\overset{N}{\sum}}x_{i}^{s}\right)  }{a(t)^{N}}\frac{sx_{i}^{s-1}%
}{a(t)^{s}}\frac{\dot{a}(t)}{a(t)}x_{i}\\
&  =0
\end{align}
The proof is completed.
\end{proof}

\begin{remark}
We notice that the novel lemma fully covers Lemma 3 in \cite{Y2} for the mass
equation (\ref{eq2}) in radial symmetry,%
\begin{equation}
\rho_{t}+\rho_{r}V+\rho V_{r}+\frac{N-1}{r}\rho V=0
\end{equation}
which showed that there exists a class of solutions
\begin{equation}
\rho(t,r)=\frac{^{f(r/a(t))}}{a(t)^{N}},\text{ }V(t,r)=\frac{\overset{\cdot
}{a}(t)}{a(t)}r,
\end{equation}
with the arbitrary function $f\geq0$ and $f\in C^{1};$ and $a(t)>0$ and
$a(t)\in C^{1}.$
\end{remark}

The proof of Theorem \ref{thm:1} is similar to the ones in \cite{DXY},
\cite{Y} and \cite{Y1}. The main idea is to put the exact solutions to check
that if they satisfy the system (\ref{pressurelessEuler}) only.

\begin{proof}
[Proof of Theorem \ref{thm:1}]From the above lemma, it is very clear to verify
that our solutions (\ref{Yuensol}) satisfy the mass equation
(\ref{pressurelessEuler})$_{1}$. Regards to the pressureless momentum equation
(\ref{pressurelessEuler})$_{2}$, we have:%
\begin{align}
&  \rho\left[  \vec{u}_{t}+(\vec{u}\cdot\nabla)\vec{u}\right] \\
&  =\rho\left[  \left(  \frac{\overset{\cdot}{a}(t)}{a(t)}\vec{x}\right)
_{t}+\frac{\overset{\cdot}{a}(t)}{a(t)}\vec{x}\cdot\nabla\frac{\overset{\cdot
}{a}(t)}{a(t)}\vec{x}\right] \\
&  =\rho\left[  \left(  \frac{\ddot{a}(t)}{a(t)}-\frac{\dot{a}^{2}(t)}%
{a^{2}(t)}\right)  \vec{x}+\frac{\dot{a}^{2}(t)}{a^{2}(t)}\vec{x}\right] \\
&  =\rho\left[  \frac{\ddot{a}(t)}{a(t)}\vec{x}\right] \\
&  =0,
\end{align}
with%
\begin{equation}
a(t)=a_{1}+a_{2}t.
\end{equation}
The proof is completed.
\end{proof}

\begin{remark}
In particular, the solutions (\ref{Yuensol}) for $s=1$,
\begin{equation}
\left\{
\begin{array}
[c]{c}%
\rho(t,\vec{x})=\frac{f\left(  \frac{1}{a(t)}\underset{i=1}{\overset{N}{\sum}%
}x_{i}^{{}}\right)  }{a(t)^{N}}\text{, }\vec{u}(t,\vec{x})=\frac
{\overset{\cdot}{a}(t)}{a(t)}\vec{x},\\
a(t)=a_{1}+a_{2}t,
\end{array}
\right.
\end{equation}
are line sources or sinks. For the physical significance of such kind of
solutions, the interested readers may refer.P.409-410 of \cite{JK} for details.
\end{remark}

On the other hand, another family of solutions are provided here.

\begin{theorem}
\label{thm:1 copy(1)}Denote $A\vec{x}:=(\frac{x_{1}}{\bar{u}_{01}},\frac
{x_{2}}{\bar{u}_{02}},.,\frac{x_{i}}{\bar{u}_{0i}}..,\frac{x_{N}}{\bar{u}%
_{0N}})$ for all $\bar{u}_{0i}\neq0$ and $A\vec{x}:=(\frac{x_{1}}{\bar{u}%
_{01}},\frac{x_{2}}{\bar{u}_{02}},.0\cdot x_{i}.,\frac{x_{N}}{\bar{u}_{0N}}),$
for some $\bar{u}_{0i}=0$. For the pressureless Euler equations
(\ref{pressurelessEuler}), there exists a family of solutions:%
\begin{equation}
\rho=f(A\vec{x}-t),\text{ }\vec{u}=\vec{u}_{0}, \label{Yuenso2}%
\end{equation}
with the arbitrary function $f\geq0$ and $f\in C^{1};$ and $\vec{u}_{0}%
=(\bar{u}_{01},\bar{u}_{02},...,\bar{u}_{0N})\in R^{N}$ is a constant vector.
\end{theorem}

It is trivial to check that the above theorem is true. We skip the proof here.

\begin{remark}
The functions (\ref{Yuensol}) and (\ref{Yuenso2}) are also the solutions to
the pressureless Navier-Stokes equations,%
\begin{equation}
\left\{
\begin{array}
[c]{rl}%
{\normalsize \rho}_{t}{\normalsize +\nabla\cdot(\rho\vec{u})} &
{\normalsize =}{\normalsize 0,}\\
\rho\left[  \vec{u}_{t}+\left(  \vec{u}\cdot\nabla\right)  \vec{u}\right]  &
{\normalsize =\mu}\nabla\left(  \nabla\cdot\vec{u}\right)  ,
\end{array}
\right.
\end{equation}
$\mu>0$ is a constant.
\end{remark}

\end{document}